\documentclass[prd,reprint,onecolumn,notitlepage,preprintnumbers, showpacs]{revtex4}

\usepackage{verbatim}
\usepackage{amsmath, multirow, amssymb, wasysym, gensymb}
\usepackage{graphicx}
\usepackage{amsfonts}
\usepackage{graphicx}
\usepackage{graphics}
\usepackage[ngerman, english]{babel}
\usepackage{float}
\usepackage{youngtab}
\usepackage{subfigure}
\usepackage[utf8]{inputenc}

\begin{document}
\title{Leptogenesis and \textbf{\textit{CP}} violation in \textbf{\textit{SU}}(5) models with lepton flavor mixing originating
from the right-handed sector
}
\author{H. P\"as}
\email{heinrich.paes@tu-dortmund.de}
\author{E. Schumacher}
\email{erik.schumacher@tu-dortmund.de}
\affiliation{Fakult\"at f\"ur Physik,
Technische Universit\"at Dortmund, 44221 Dortmund,
Germany}
\preprint{DO-TH 14/01}

\begin{abstract}
\noindent
We discuss neutrino masses and mixing in the context of seesaw type I models with three right-handed Majorana neutrinos and an approximately diagonal Dirac sector. This ansatz is motivated by the idea that the flavor structure in the right-handed Majorana masses is responsible for the large mixing angles, whereas the small mixing angle $\theta_{13}$ originates from the Dirac Yukawa couplings in analogy to the quark sector. To obtain $\theta_{13} \approx 0.15$ we study a possible $SU(5)$ grand unified theory realization with a $U(1) \times \mathbb{Z}'_2 \times \mathbb{Z}''_2 \times \mathbb{Z}'''_2$ flavor symmetry and include a complex perturbation parameter in the Dirac mass matrix. The consequences for $CP$ violating phases and effects on leptogenesis are investigated.
\end{abstract}
\maketitle
\section{Introduction}
\label{s:intro}
Neutrinos are special in several respects: First, they are much lighter than the charged leptons and quarks and as they do not carry any unbroken quantum number they can be Majorana particles. These properties are exploited in the seesaw mechanism where the heavy Majorana
masses of right-handed neutrinos drive the effective masses of left-handed neutrinos down to or below the eV scale.
Moreover, the mismatch between neutrino and charged lepton mixing parametrized in the Pontecorvo-Maki-Nakagawa-Sakata (PMNS) matrix exhibits large or even maximal
mixing angles, in stark contrast to the small Cabibbo-Kobayashi-Maskawa (CKM) mixing in the quark sector. After recent data from reactor experiments have revealed a nonvanishing neutrino mixing angle $\theta_{13}$ and thus rule out exact tribimaximal (TBM) mixing \cite{Ahn:2012nd,An:2012eh,Abe:2013sxa}, many attempts have been made to explain the finite $\theta_{13}$. While the idea of anarchy is becoming more attractive, the most popular approaches are still based on discrete symmetries, as summarized, e.g., in \cite{Antusch:2013ti} or \cite{King:2013eh}. Models that intend to explain the leptonic mixing pattern can be constructed using large symmetry groups such as $\Delta(96)$ or $D_{14}$. At the expense of predictivity, these models contain several free parameters to account for all the physical observables. Alternatively, models can be based on smaller symmetry groups, e.g., $A_4$ or $S_4$, which yield specific patterns such as the TBM or golden ratio mixing and include perturbations in order to generate the necessary deviations from these structures. 

In this paper we analyze the possibility that also the large lepton mixing arises from the right-handed Majorana sector. To this end we study a generic type I seesaw model with three heavy right-handed neutrinos:
\begin{align}
\label{e:seesaw}
m_{\nu}= m_D^T \, M_R^{-1}\,m_D,
\end{align}
where $m_D$ denotes the Dirac mass matrix and $M_R$ the matrix of the right-handed neutrinos. In a previous publication \cite{Leser:2011fz}
the Dirac matrix has been assumed to be diagonal, and it has been shown that this can give rise to TBM. We adjust this pattern by perturbing the Dirac mass matrix with a complex
parameter in order to accommodate a finite $\theta_{13}$. The philosophy behind this ansatz is to have the small mixing originate from the Dirac
sector, in analogy to the small CKM mixing in the quark sector, while the Majorana property of the right-handed neutrinos is responsible for 
the large mixing angles $\theta_{23}$ and $\theta_{12}$. The structure of the Dirac sector is motivated by an $SU(5)$ Grand Unified Theory (GUT) 
embedding, which accounts not only for leptonic mixing but also for observables of the quark sector. Related works exist, where the Majorana mass terms of the neutrino sector account for the large leptonic mixing angles, e.g., \cite{Alonso:2013mca,Alonso:2012fy}

The paper is organized as follows: We start by describing the outline of the model in Sec. \ref{s:model} proposing a possible realization in an
$SU(5)$ GUT. In Sec. \ref{s:cp} we briefly summarize the methods used to analyze $CP$ violation in our model and then focus on leptogenesis in Sec. \ref{s:lepto}. Sec. \ref{s:num} deals with the numerical analysis and with the implications for neutrinoless double-beta decay ($0\nu\beta\beta$). Conclusions are drawn in Sec. \ref{s:con}.

\section{Outline of the model}
\label{s:model}
If we consider $m_D$ and the mass matrix of the charged leptons to be diagonal, all mixing in the neutrino sector originates from $M_R$. An explicit example can be constructed by using an $SU(5)$ GUT with  $U(1) \times \mathbb{Z}'_2 \times \mathbb{Z}''_2 \times \mathbb{Z}'''_2$ 
flavor symmetry in close analogy to the model published in 
\cite{McKeen:2007ry}. All standard model particles, including the right-handed neutrinos, are accommodated in the 
\textbf{10} $\oplus$ \textbf{5} $\oplus$ \textbf{1} multiplets of $SU(5)$. Assigning appropriate quantum numbers under these symmetry groups yields the quark mixing matrix in the first approximation, where the off-diagonal elements are suppressed by a Froggatt-Nielsen (FN) mechanism \cite{Froggatt:1978nt}. In this framework the fermion masses are generated through couplings to additional scalar fields $\eta_i$ with universal vacuum expectation values (VEVs) $u$. These VEVs are suppressed by a large messenger scale $\Lambda$ such that $\frac{u}{\Lambda} \approx \lambda =0.22$, consequently leading to suppression factors in the fermion mass term $\propto \lambda^n$, where $n$ denotes the sum of the fermion field charges under the corresponding flavor symmetries. It has been shown, e.g., in \cite{Altarelli:2012ia,Campos:2014lla} that the combination of an $SU(5)$ with a $U(1)_{\text{FN}}$ symmetry can give rise to maximal mixing in the lepton sector due to specific $U(1)_{\text{FN}}$ charge assignments.

The matrix structures generated by the FN mechanism must be consistent with the following approximate mass relations:
\begin{align}
 m_u:m_c:m_t\,\approx\,\lambda^8:\lambda^4:1\,,\qquad
 m_d:m_s:m_b\,\approx\,\lambda^4:\lambda^2:1\,,\qquad
 m_e:m_{\mu}:m_{\tau}\,\approx\,\lambda^4:\lambda^2:1\,.
\end{align}
Following Ref. \cite{McKeen:2007ry}, assigning the following $U$(1) charges to the $SU(5)$ multiplets, and (-1) for the flavon fields, leads to the desired structures of the Yukawa matrices in the quark sector,
\begin{align}
\label{e:char}
\textbf{10} \quad &: \quad I:\,4\, \quad II:\,2\, \quad III:\,0, \\
\textbf{5} \quad &: \quad I:\,3\, \quad II:\,3\, \quad III:\,3,
\end{align}
where $I,II$ and $III$ specify the family number. 
The powers of the suppression factors $\lambda$ entering the Lagrangian depend on the $\mathbb{Z}_2$ charges $\rho_1, \rho_2$ and $\rho_3$
of the respective flavons $\eta', \eta''$, and $\eta'''$. The sum rules for cyclic groups $\mathbb{Z}_2$ are
\begin{align}
\label{e:sumq}
 \rho_i + \rho_j = (\rho_i + \rho_j)~\text{mod}~2,
\end{align}
where $\rho_{i,j}$ again are the $\mathbb{Z}_n$ charges of the interacting fermions $\psi_{i,j}$. This results in
\begin{align}
 \textbf{10}_i \otimes \textbf{10}_j \quad &: \quad 
Y_u \propto \left(
\begin{array}{ccc}
\lambda^8 & \lambda^{6+\rho_1+\rho_2} & \lambda^{4+\rho_1+\rho_3} \\
\lambda^{6+\rho_1+\rho_2} & \lambda^4 & \lambda^{2+\rho_3+\rho_2} \\
\lambda^{4+\rho_1+\rho_3} & \lambda^{2+\rho_3+\rho_2} & 1
\end{array}\right),\\
\textbf{10}_i \otimes \overline{\textbf{5}}_j \quad &: \quad 
Y_d \propto \left(
\begin{array}{ccc}
\lambda^7 & \lambda^{7+\rho_1+\rho_2} & \lambda^{7+\rho_1+\rho_3} \\
\lambda^{5+\rho_1+\rho_2} & \lambda^5 & \lambda^{5+\rho_3+\rho_2} \\
\lambda^{3+\rho_1+\rho_3} & \lambda^{3+\rho_3+\rho_2} & \lambda^3
\end{array}\right).
\end{align}
Since the Yukawa matrix of the charged leptons $Y_l \sim \overline{\textbf{5}}_i \otimes \textbf{10}_j \sim Y_d^T$ is hierarchical as well, large lepton mixing must be a consequence of the specific structure of the neutrino sector. To establish the structure of the latter, let us denote the $U$(1) charges of the right-handed neutrinos $N^R_{1,2,3}$ in the $SU$(5) singlet \textbf{1} as $e_1,e_2$, and $e_3$, respectively. According to Ref. \cite{McKeen:2007ry} the right-handed neutrinos carry no $\mathbb{Z}_2$ charges as the seesaw scale is lower than the messenger scale $\Lambda$ at which the flavons $\eta_i$ receive their VEVs $u$. The relevant matrix structures in the neutrino sector arise from the following products:
\begin{align}
\overline{\textbf{5}}_i \otimes \textbf{1}_j \quad &: \quad 
Y^D \propto \left(
\begin{array}{ccc}
\lambda^{3+e_1} & \lambda^{3+e_2+\rho_1} & \lambda^{3+e_3+\rho_1} \\
\lambda^{3+e_1+\rho_2} & \lambda^{3+e_2} & \lambda^{3+e_3+\rho_2} \\
\lambda^{3+e_1+\rho_3} & \lambda^{3+e_2+\rho_3} & \lambda^{3+e_3}
\end{array}\right), \\
\textbf{1}_i \otimes \textbf{1}_j \quad &: \quad 
Y_R \propto \left(
\begin{array}{ccc}
\lambda^{2e_1} & \lambda^{e_1+e_2} & \lambda^{e_1+e_3} \\
\lambda^{e_2+e_1} & \lambda^{2e_2} & \lambda^{e_2+e_3} \\
\lambda^{e_3+e_1} & \lambda^{e_3+e_2} & \lambda^{2e_3}
\end{array}\right).
\end{align}
By choosing 
\begin{align}
\label{e:charges}
 e_3=0;\quad e_1=e_2=1;\quad \rho_1=\rho_2=\rho_3=1\,,
\end{align}
we can motivate a hierarchical structure of the Dirac matrix $Y^D$ with two perturbation parameters assigned to the elements $Y^D_{12}$ and $Y^D_{23}$: 
\begin{align}
\label{e:yd}
Y^D \propto \lambda^3 \cdot \left(
\begin{array}{ccc}
\lambda & \lambda^{2} & \lambda \\
\lambda^{2} & \lambda & \lambda \\
\lambda^{2} & \lambda^{2} & 1
\end{array}\right),\qquad 
Y_R \propto \left(
\begin{array}{ccc}
\lambda^{2} & \lambda^{2} & \lambda \\
\lambda^{2} & \lambda^{2} & \lambda \\
\lambda & \lambda & 1
\end{array}\right).
\end{align}
Following our charge assignments from Eqs. (\ref{e:charges}) and (\ref{e:yd}) we choose a specific Dirac mass pattern for our further analysis,
\begin{align}
\label{e:cg}
m_D = \left( \begin{array}{ccc}
\lambda & 0 & \epsilon \\
0 & \lambda & \gamma \\
0 & 0 & 1
\end{array}\right)\cdot v,
\end{align}
with $\lambda \approx 0.22$ and $v=\frac{246}{\sqrt{2}}\,$GeV, where $\epsilon$ generates a nonzero $\theta_{13}$ value and $\gamma$ cancels the effects of $\epsilon$ on the large mixing angles $\theta_{12}$ and $\theta_{23}$. In analogy to the quark sector we assume that all complex phases except for one can be absorbed into the interacting lepton fields. A single complex phase $\phi$ in $\epsilon=|\epsilon|e^{i\phi}$ remains, whereas $\gamma \in \mathbb{R}$.

The procedure to accommodate the experimentally determined mixing angles can be summarized in three steps. 
It has been demonstrated in Ref. \cite{Leser:2011fz} that a diagonal Dirac sector can give rise to TBM mixing. Therefore by adopting
\begin{align}
\label{e:md}
 m_D = \text{diag}(\lambda, \lambda, 1)\cdot v
\end{align}
we first determine the right-handed mass matrix $M_R$ to yield exact TBM mixing with $\theta_{12}$ and $\theta_{23}$ in the allowed 3$\sigma$ ranges \cite{GonzalezGarcia:2012sz},
\begin{align}
\label{e:bounds}
\theta_{12} \in \left[0.543,0.626\right],\quad
\theta_{23} \in \left[0.625,0.956\right],\quad
\theta_{13} \in \left[0.125,0.173\right].
\end{align}
We can derive an analytical expression for $M_R^{-1}$ from the condition that $m_{\nu}$ must be diagonalizable with $U_{\text{TBM}}$:
\begin{align}
\label{e:mrinv}
 M_R^{-1}&=(m_D^T)^{-1}\,U_{\text{TBM}}\,m'_{\nu}\,U_{\text{TBM}}^T\,m_D^{-1}\\
&= \frac{1}{3v^2}\,\left(
\begin{array}{ccc}
 \frac{2 m_1+m_2}{\lambda^2} & \frac{-m_1+m_2}{\lambda^2} & \frac{-m_1+m_2}{\lambda} \\
 \frac{-m_1+m_2}{\lambda^2} & \frac{m_1+2 m_2+3 m_3}{2 \lambda^2} & \frac{m_1+2 m_2-3 m_3}{2 \lambda} \\
 \frac{-m_1+m_2}{\lambda} & \frac{m_1+2 m_2-3 m_3}{2 \lambda} & \frac{m_1+2 m_2+3 m_3}{2}
\end{array}
\right) \quad \text{and} \\
\label{e:mr}
 M_R&= \frac{v^2}{3}\,\left(
\begin{array}{ccc}
 \frac{\lambda^2 \left(m_1+2 m_2\right)}{m_1 m_2} & \frac{\lambda^2 \left(m_1-m_2\right)}{m_1 m_2} & \frac{\lambda \left(m_1-m_2\right)}{m_1 m_2} \\
 \frac{\lambda^2 \left(m_1-m_2\right)}{m_1 m_2} & \frac{1}{2} \lambda^2 \left(\frac{1}{m_1}+\frac{2}{m_2}+\frac{3}{m_3}\right) & \frac{1}{2} \lambda \left(\frac{1}{m_1}+\frac{2}{m_2}-\frac{3}{m_3}\right) \\
 \frac{\lambda \left(m_1-m_2\right)}{m_1 m_2} & \frac{1}{2} \lambda \left(\frac{1}{m_1}+\frac{2}{m_2}-\frac{3}{m_3}\right) & \frac{1}{2} \left(\frac{1}{m_1}+\frac{2}{m_2}+\frac{3}{m_3}\right)
\end{array}
\right)
\end{align}
where $m'_{\nu} = \text{diag}(m_1,m_2,m_3)$ serves as an input parameter. The matrix $M_R$ given in Eq. (\ref{e:mr}) is in good agreement with $Y_R$ of Eq. (\ref{e:yd}). The initial values for the light neutrino masses $m_{i}$ in $m'_{\nu}$ are selected according to the experimental bounds on $\Delta m_{12}^2$ and $\Delta m_{23}^2$ \cite{GonzalezGarcia:2010er}
\begin{align}
\label{e:deltam}
\Delta m_{12}^2 = 7.59 \times 10^{-5}\,(\text{eV})^2, \qquad \Delta m_{23}^2 = \begin{cases} +2.46 \times 10^{-3}\,(\text{eV})^2~\text{(NH)} \\
											    -2.37 \times 10^{-3}\,(\text{eV})^2~\text{(IH)}\end{cases}.
\end{align}
The model exhibits different behavior in the various mass regimes, where the following cases are considered:
\begin{enumerate}
\item Degenerate masses (deg): $m_1 \approx m_2 \approx m_3 > 0$,
\item Normal hierarchy (NH): $m_3 \gg m_2 \approx m_1 \approx 0$,
\item Inverse hierarchy (IH): $m_1 \approx m_2 \gg m_3 \approx 0$.
\end{enumerate}

In the second step we include the perturbation parameters $\epsilon$ and $\gamma$ in the Dirac mass matrix $m_D$ according to Eq. (\ref{e:cg}) to produce $\theta_{13} \approx 0.15 = \theta_{13}^{\text{Exp}}$, while $M_R$ remains unchanged. This guarantees that $\theta_{13}$ is solely affected by Dirac couplings. 

The parameters $|\epsilon|, \phi$ and $\gamma$ then are fitted to the current experimental $3\sigma$ bounds on the mixing angles $\theta_{ij}$ given in Eq. (\ref{e:bounds}) for neutrino mass eigenstates $m_i \in [10^{-3},10^{-1}]\,$eV. This mass region includes the NH and IH scenarios as well as the degenerate mass regime being consistent with cosmological bounds on the neutrino masses \cite{Ade:2013ktc} and a successful leptogenesis scenario \cite{DiBari:2004en}.
The bounds on the mixing angles constrain the parameter space of $\epsilon$ to small regions, enabling predictions on the observable $CP$ phases. Assuming that leptogenesis successfully generates the correct baryon asymmetry of the Universe, additional constraints arise that confine these intervals even further.

\section{$CP$ violation}
\label{s:cp}
The PMNS matrix depends on four parameters: three mixing angles $\theta_{ij}$ and one $CP$ violating phase $\delta$. In the case of three additional right-handed neutrinos the mixing matrix includes two supplementary Majorana phases, leading to the conventional parametrization of the PMNS matrix
\begin{align}
 \label{e:conv}
U_{\text{PMNS}}= \left(
\begin{array}{ccc}
 c_{12} c_{13} & s_{12} c_{13} & s_{13} e^{-i\delta} \\
 -s_{12}c_{23} -c_{12}s_{23}s_{13}e^{i\delta} & c_{12}c_{23}-s_{12}s_{23}s_{13}e^{i\delta}& s_{23} c_{13} \\
  s_{12}s_{23} -c_{12}c_{23}s_{13}e^{i\delta} & -c_{12}s_{23}-s_{12}c_{23}s_{13}e^{i\delta} & c_{23} c_{13} 
\end{array}
\right)\cdot P \,, 
\end{align}
where $P=\text{diag}(e^{i\alpha},e^{i\beta},1)$ contains the Majorana phases $\alpha$ and $\beta$, and $s_{ij}$ and $c_{ij}$ are abbreviations for $\sin \theta_{ij}$ and $\cos \theta_{ij}$, respectively.

In our model the complex perturbation parameter $\epsilon$ provides the source of $CP$ violation in the neutrino sector and leads to $CP$ violating phases in $U$. The mixing matrix can be extracted from $m'_{\nu} = U^T\, m_{\nu}\, U$ using a Takagi decomposition, which is applicable for complex symmetric matrices \cite{Hahn:2006hr}. The real diagonal matrix $m'_{\nu}$ contains the non-negative square roots of the eigenvalues of $m_{\nu} m_{\nu}^{\dagger}$. However, the mixing matrix we receive from our numerical analysis does not resemble the usual convention of the PMNS matrix given in Eq. (\ref{e:conv}). We can extract the mixing angles $\theta_{ij}$ and $CP$ phases $\delta$, $\alpha$ and $\beta$ using rephasing invariants and properties of the unitarity triangles of $U$, previously explored in \cite{Branco:2011zb,Branco:1986gr}. By defining "Majorana-type`` phases $\zeta_i$ and $\xi_i$, $i \in \left(1,2,3\right)$,
\begin{align}
\zeta_1 &\equiv \text{Arg}(U_{e1} U_{e2}^{\ast}), \quad \zeta_2 \equiv \text{Arg}(U_{\mu1} U_{\mu2}^{\ast}), \quad \zeta_3 \equiv \text{Arg}(U_{\tau1} U_{\tau2}^{\ast}), \\
\xi_1 &\equiv \text{Arg}(U_{e1} U_{e3}^{\ast}), \quad \xi_2 \equiv \text{Arg}(U_{\mu1} U_{\mu3}^{\ast}), \quad \xi_3 \equiv \text{Arg}(U_{\tau1} U_{\tau3}^{\ast})\,,
\end{align}
we can express the mixing angles and $\delta$ according to \cite{Branco:2008ai} as a function of $\zeta_i$ and $\xi_i$:
\begin{align}
\tan^2   \theta_{12} &= \frac{|\sin(\xi_1 -\xi_2)| 
|\sin(-\zeta_2 + \xi_2 + \zeta_3 -\xi_3)| 
|\sin(\xi_1 -\xi_3)| }
{|\sin(-\zeta_1 + \xi_1 + \zeta_2 -\xi_2)|
 |\sin(\xi_2 -\xi_3)| |\sin(-\zeta_1 + \xi_1 + \zeta_3 -\xi_3)|},
\label{e:th12} \\
\tan^2   \theta_{23} &= \frac{|\sin(\xi_1 -\xi_3)| 
|\sin(-\zeta_1 + \xi_1 + \zeta_3 -\xi_3)| 
|\sin(\zeta_1 -\zeta_2)| }
{|\sin(-\zeta_1 + \xi_1 + \zeta_2 -\xi_2)|
 |\sin(\xi_1 -\xi_2)| |\sin(\zeta_1 - \zeta_3 )|},
\label{e:th23} \\
\label{e:th13}
\tan^2   \theta_{13} &=  \frac{|\sin(\xi_2 -\xi_3)| 
|\sin(\zeta_1 -\zeta_3)|
|\sin(\zeta_1 -\zeta_2)| }
{|\sin(\xi_1 -\xi_3)|
 |\sin(\xi_1 -\xi_2)| |\sin(\zeta_2 - \zeta_3 )|} \cdot \sin^2 \theta_{12}, \\
 \frac{1}{8} |\sin \delta| &= \frac{|\cos \theta_{12} \cos \theta_{13} \sin \theta_{13}|^2 |\sin(\xi_1 - \xi_3)| |\sin(\xi_1 - \xi_2)|}{|\sin (2\theta_{12}) \sin (2\theta_{13}) \sin (2\theta_{23}) \cos \theta_{13} | |\sin(\xi_2 - \xi_3)|}.
\end{align}
These formulas are valid regardless of the parametrization of the matrix $U$. To extract the Majorana phases $\alpha$ and $\beta$ we recall that they rotate the Majorana unitarity triangles in the complex plane. Hence, one can conclude that expressions for the phases $\alpha$ and $\beta$ result from (with $U_{\text{PMNS},ij}:=U_{Pij}$)
\begin{align}
\text{Arg}\left(\frac{U_{P11} U_{P13}^{\ast}}{U_{11} U_{13}^{\ast}}\right) = \alpha,\quad
\text{Arg}\left(\frac{U_{P12} U_{P13}^{\ast}}{U_{12} U_{13}^{\ast}}\right) = \beta,\quad
\text{Arg}\left(\frac{U_{P11} U_{P12}^{\ast}}{U_{11} U_{12}^{\ast}}\right) = \alpha-\beta.
\end{align}

\section{Leptogenesis}
\label{s:lepto}
Leptogenesis \cite{Fukugita:2002hu} explains the present matter asymmetry by assuming that a lepton asymmetry in the early Universe is converted into a baryon asymmetry through $B+L$ violating sphaleron processes. Leptogenesis is closely connected to the seesaw mechanism as it relies on the existence of heavy right-handed Majorana neutrinos $N_i$ decaying into leptons $l_{\alpha}$ and scalar fields $\phi$ via $N_i \rightarrow l_{\alpha} + \phi$. Observations of the cosmic microwave background allow for an estimate of the baryon asymmetry \cite{Larson:2010gs}
\begin{align}
\label{e:cmb}
Y_{\Delta \text{B}}^{\text{CMB}} = \frac{n_B-n_{\overline{B}}}{s} = (8.79 \pm 0.44)\cdot 10^{-11}\,,
\end{align}
where $s$ is the entropy density of the universe and $n_{B,\overline{B}}$ denote the abundances of baryons and antibaryons, respectively. The baryon asymmetry can be calculated in terms of the $CP$ asymmetry $\sigma_{i\alpha}$ and the efficiency factor $\kappa_i$ \cite{Fong:2013wr},
\begin{align}
\label{e:baryo}
Y_{\Delta \text{B}} = \sigma_{i\alpha}\,\kappa_i\,10^{-3}. 
\end{align}
The efficiency factor $\kappa_i$ measures the effect of washout processes depending on the ''washout regime''. For a general system it is given by the solution of the Boltzmann equations for leptogenesis, which characterize the competition of production and washout of the $N_i$'s. In this paper the following approximations are used, as described in Ref.~\cite{Fong:2013wr}:
\begin{align}
\label{e:wash}
&\widetilde{m} < m_{\ast} ~\cap~ m_i < m_{\ast}:\quad \kappa_i \approx \frac{m_i \widetilde{m}}{m_{\ast}^2}\quad \text{weak washout}\\ 
\label{e:int} &\widetilde{m} > m_{\ast} ~\cap~ m_i < m_{\ast}:\quad \kappa_i \approx \frac{m_i}{m_{\ast}}\quad \text{intermediate washout}\\
\label{e:strong} &\widetilde{m} > m_{\ast} ~\cap~ m_i > m_{\ast}:\quad \kappa_i \approx \frac{m_{\ast}}{m_i}\quad \text{strong washout}
\end{align}
with $m_{\ast} \approx 1.1 \times 10^{-3}\,$eV and the sum of the light neutrino masses $\widetilde{m}=\sum_i m_i$.

The $CP$ asymmetry $\sigma_{i\alpha}$ generated by a heavy right-handed neutrino $N_i$ that decays into a lepton with flavor $\alpha$ reads explicitly \cite{Fong:2013wr}
\begin{align}
\label{e:cp}
\sigma_{i\alpha} &\equiv \frac{1}{8\pi} \frac{1}{(Y^{D \dagger}Y^D)_{ii}} \sum_{j \neq i} \left\{\text{Im} \left[ (Y^{D\dagger}Y^D)_{ji} Y_{\alpha i} Y_{\alpha j}^{D \ast} \right]g\left(\frac{M_j^2}{M_i^2}\right)+ \text{Im} \left[ (Y^{D \dagger}Y^D)_{ij} Y^D_{\alpha i} Y_{\alpha j}^{D \ast} \right]\left(\frac{M_i^2}{M_i^2-M_j^2}\right) \right\}.
\end{align}
The Yukawa couplings matrix $Y^D$ is related to the Dirac matrix by $Y^D=\frac{m_D}{v}$. Because of the Majorana nature of the $N_i$, contributions to the $CP$ asymmetry arise only from higher-order interferences of their decays. The loop corrections are included in the function $g(x)$
\begin{align}
g(x)&=\sqrt{x} \left[\frac{1}{1-x} + 1 -(1+x)\ln\left(1+\frac{1}{x}\right)\right].
\end{align}
In a model with three additional right-handed neutrinos, the neutralizing effect of the $N_{2,3}$ on the asymmetry generated in the decay of $N_1$ can be neglected if $M_{2,3} > T_{\text{reh}}$, the reheating temperature, and $M_1 \ll M_{2,3}$. In our approximation the second term in Eq. (\ref{e:cp}) vanishes due to zeros in the pattern of $Y^{D}$.

Model independent limits on neutrino masses can be inferred from the upper bound of the $CP$ asymmetry $\sigma_{i\alpha}$. As stated in Sec. \ref{s:model}, in the single lepton flavor approximation with hierarchical right-handed neutrino masses $M_i$ a successful leptogenesis mechanism is confined to the mass region $10^{-3}\,\text{eV}<\,m_i<\,0.1\,$eV \cite{DiBari:2004en}. We will therefore focus our attention on this interval.

Because of the specific structure of $m_D$ given in Eq. (\ref{e:cg}), we obtain a small number of contributions to the $CP$ asymmetry. With
\begin{align}
Y^D=\left(
\begin{array}{ccc}
\lambda & 0 & |\epsilon|e^{i\phi}  \\
 0 & \lambda & \gamma \\
 0 & 0 & 1
\end{array}
\right),\quad Y^{D\ast}=\left(
\begin{array}{ccc}
\lambda & 0 & |\epsilon|e^{-i\phi}  \\
 0 & \lambda & \gamma \\
 0 & 0 & 1
\end{array}
\right),\quad \text{and} \quad Y^{\dagger}Y = \left(
\begin{array}{ccc}
\lambda^2 & 0 & |\epsilon|e^{i\phi} \lambda \\
0 & \lambda^2 & \gamma \lambda \\
|\epsilon|e^{-i\phi} \lambda & \gamma \lambda & 1
\end{array}\right),
\end{align}
we receive nonvanishing contributions to $\sigma_{i\alpha}$ only if $\alpha = 1$: 
\begin{align}
 \label{e:ybgam}
\sigma_{1}=\sum_{i} \sigma_{i1} = \frac{|\epsilon|^2 \sin (2\phi)}{8\pi} \left(\lambda^2\cdot g\left(\frac{M_1^2}{M_3^2}\right) - g\left(\frac{M_3^2}{M_1^2}\right)\right),
\end{align}
resulting in a $|\epsilon|^2 \sin (2\phi)$ dependence of $Y_{\Delta\text{B}}$. Note that the parameter $\gamma$ does not affect the generated $CP$ asymmetry.

\section{Numerical analysis}
\label{s:num}
For the numerical analysis we use the following ordering schemes:
\begin{align}
\label{e:norm}
\text{NH:}\quad m_1 = m_0,\quad m_2 = \sqrt{m_0^2 + \Delta m_{12}^2},\quad m_3 = \sqrt{m_0^2 + \Delta m_{12}^2 + \Delta m_{23}^2}\,,\\
\text{IH:}\quad m_3 = m_0,\quad m_2 = \sqrt{m_0^2 + \Delta m_{12}^2 + \Delta m_{23}^2 },\quad m_1 = \sqrt{m_0^2  + \Delta m_{23}^2}\,,
\end{align}
where $m_0 \in [10^{-3},10^{-1}]\,$eV. In the regions of larger masses, where $m_i \gg \sqrt{\Delta m_{12}^2},\,\sqrt{\Delta m_{23}^2}$ the mass eigenstates are degenerate. To explain the small neutrino masses through heavy right-handed neutrinos, which are compatible with the leptogenesis mechanism described in Sec. \ref{s:lepto}, the Dirac masses have to be at the GeV scale. As explained in Sec. \ref{s:intro}, we assume TBM mixing with a diagonal Dirac sector in order to determine $M_R$ from Eq. (\ref{e:mrinv}).

In the cases of $m_0 = 10^{-3}\,$eV, where the NH and IH scenarios are relevant, and $m_0 = 0.1\,$eV (degenerate masses) the right-handed mass eigenstates are given by
\begin{align}
\text{NH:} &\qquad M_{\text{R},1} = 4.934 \cdot 10^{13}\,\text{GeV}, \quad M_{\text{R},2} = 3.942 \cdot 10^{14}\,\text{GeV}, \quad M_{\text{R},3} = 7.311 \cdot 10^{15}\,\text{GeV},\\
\text{IH:} &\qquad M_{\text{R},1} = 2.890 \cdot 10^{13}\,\text{GeV}, \quad M_{\text{R},2} = 5.452\cdot 10^{13}\,\text{GeV}, \quad M_{\text{R},3} = 1.595 \cdot 10^{16}\,\text{GeV},\\
\text{deg:} &\qquad M_{\text{R},1} = 1.365 \cdot 10^{13}\,\text{GeV}, \quad M_{\text{R},2} = 1.447 \cdot 10^{13}\,\text{GeV}, \quad M_{\text{R},3} = 2.830 \cdot 10^{14}\,\text{GeV},
\end{align}
revealing a hierarchy among the the heavy right-handed neutrino masses, fulfilling the conditions for our approximations in leptogenesis.

The parameter spaces of the perturbation parameters $|\epsilon|$, $\phi$, and $\gamma$ are scanned for combinations that reproduce the neutrino mixing angles $\theta_{ij}$ within the current $3\sigma$ bounds, see Eq. (\ref{e:bounds}). Assuming that leptogenesis successfully generates the correct baryon asymmetry of the Universe, given in Eq. (\ref{e:cmb}), we can confine these regions even further to make precise predictions on physical observables. 

The comparison of low-scale experimental data on neutrino parameters with calculations carried out at the GUT scale requires taking into account the renormalization group (RG) evolution of the leptonic mixing parameters. The RG running effects on the mixing parameters have been considered in various publications \cite{Chankowski:1993tx,Antusch:2005gp,Cooper:2011rh} and can be significant especially in the case of a degenerate mass spectrum. The leptonic mixing angles are expected to run faster than their quark equivalents for they are larger and the neutrino mass differences are particularly tiny. The generic enhancement for the evolution of the angles can be estimated analytically, which has been done, for instance, in \cite{Antusch:2005gp}. 
For our numerical analysis of the degenerate mass regime we calculate the light neutrino masses at the seesaw scale using Eq. (\ref{e:seesaw}) and run the matrix $m_{\nu}$ down to the electroweak scale before determining the mixing parameters according to Sec. \ref{s:cp}. This allows for a better comparison with experimental data and a more robust analysis. The effects of RG running are implemented using REAP 1.8.4, generously provided by \cite{Antusch:2005gp}. 

\begin{figure}[H]
	\centering
		\subfigure[$Y_{\Delta\text{B}}(|\epsilon|,\phi),\,$NH]{\includegraphics[width=0.49\textwidth]{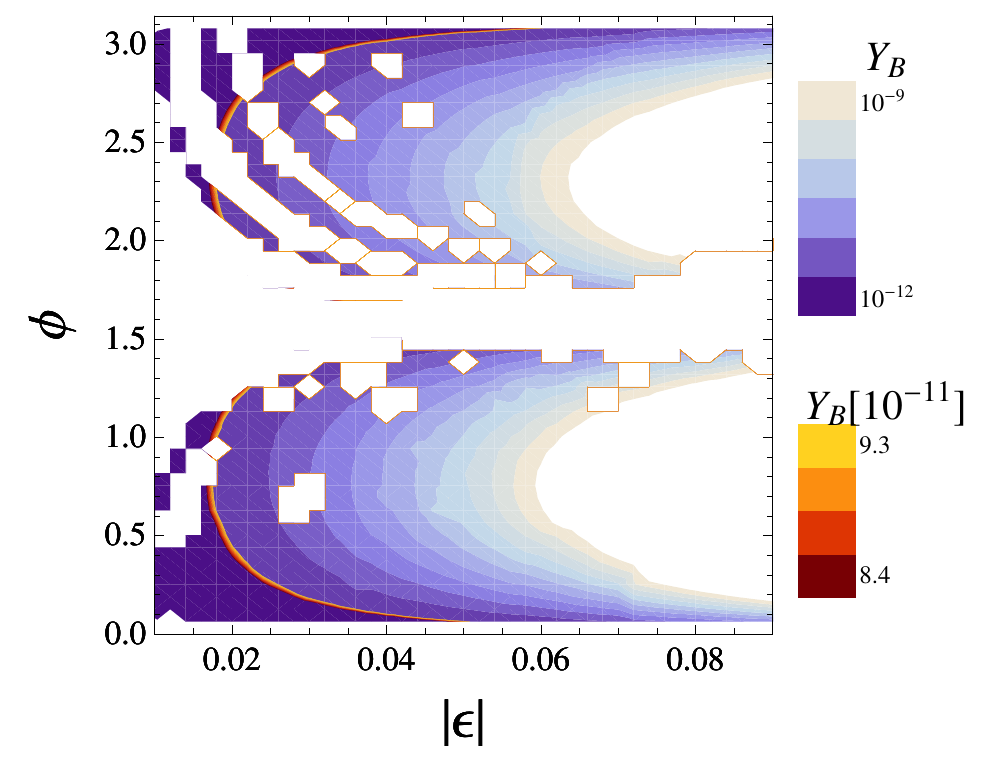}}
		\subfigure[$Y_{\Delta\text{B}}(|\epsilon|,\phi),\,$IH]{\includegraphics[width=0.49\textwidth]{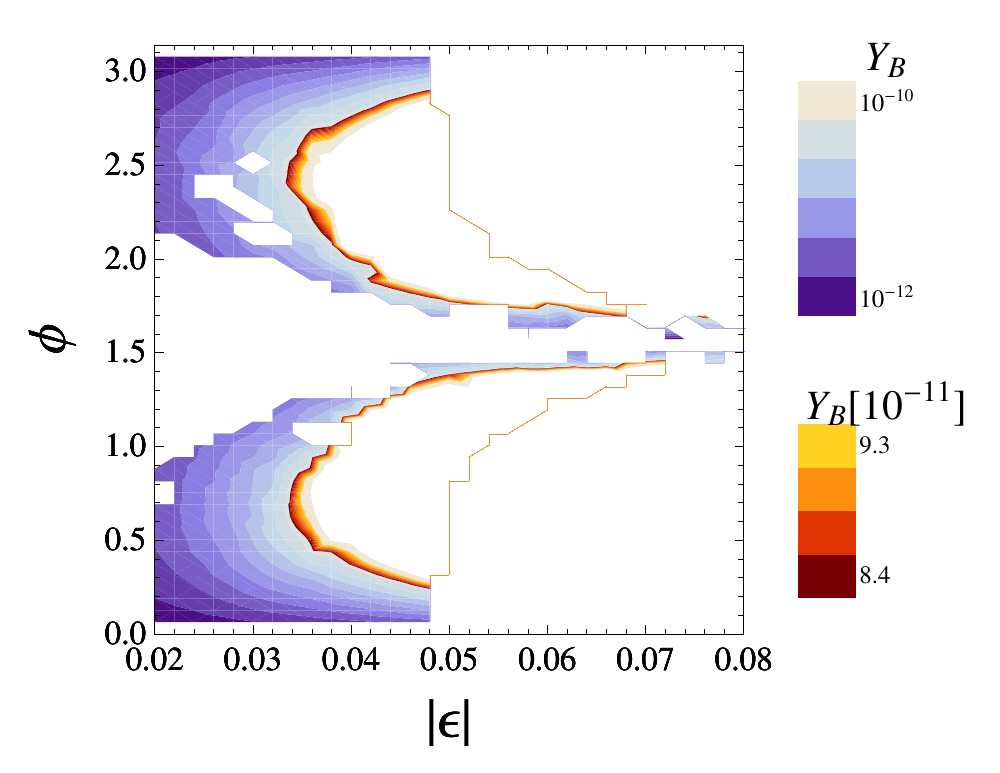}}
	\caption{(a): Parameter space of $|\epsilon|$ and $\phi$ compatible with the 3$\sigma$ ranges of the $\theta_{ij}$ in the case of NH with $m_0 \in \left[0.00, 0.05\right]\,\text{eV}$. The white areas did not yield models compatible with the experimental bounds. The (red) line overlying the contour plot marks parameter values that successfully generate the baryon asymmetry $Y_{\Delta\text{B}} = (8.79 \pm 0.44)\times 10^{-11}$ in leptogenesis. (b) Parameter space in the IH case.}
	\label{fig:bar}
\end{figure}

Regarding leptogenesis, Eq. (\ref{e:strong}) is used to compute the efficiency factor $\kappa_i$ since $m_0 > 10^{-3} > m^{\ast}$. The resulting parameter spaces complying with Eq. (\ref{e:cmb}) are listed in Table \ref{tab:lep}, where $\delta, \alpha, \beta$ denote the Dirac and Majorana $CP$ phases, respectively, corresponding to the parameter regions of $|\epsilon|$, $\phi$, and $\gamma$. The allowed combinations of $|\epsilon|$ and $\phi$ as well as the generated baryon asymmetry are depicted in Fig. \ref{fig:bar}. The blue areas denote parameter values that account for neutrino mixing, while the red color indicates combinations that also lead to successful leptogenesis. 

\begin{table}[H]
	\centering
		\begin{tabular}{|r|c|c|c|c|}\hline
				& $m_0$ &$|\epsilon|$ &$\phi$ &$\gamma$\\\hline
			NH& $\left[0.018, 0.050\right]$&$\left[0.018, 0.060\right]$ &$\left[0.063, 0.565\right] \cup \left[2.639, 3.079\right]$ &$\left[0.02, 0.20\right]$\\ 
			IH& $\left[0.002, 0.044\right]$& $\left[0.034, 0.076\right]$&$\left[0.251, 1.445\right] \cup \left[1.696, 2.890\right]$ & $\left[-0.38, 0.00\right]$ \\ \hline
			  & & $\delta$ &$\alpha$ &$\beta$\\\hline
			NH& & $\left[0.016, 0.083\right]$ &$\left[0.039, 0.192\right] \cup \left[3.118, 3.133\right]$ & $\left[0.003, 0.157\right]$\\ 
			IH& & $\left[0.205, 1.293\right]$&$\left[0.448, 3.112\right]$ &$\left[0.009, 0.067\right] \cup \left[0.402, 2.404\right]$\\ \hline
		\end{tabular}
	\caption{Parameter spaces of the fit parameters $|\epsilon|$, $\phi$ and $\gamma$ for successful leptogenesis and the $CP$ phases $\delta,\alpha$, and $\beta$ corresponding to these intervals. }
	\label{tab:lep}
\end{table}

In a previous publication \cite{Leser:2011fz} it was found that models with a diagonal Dirac sector and TBM mixing favor very light neutrino masses. Although in principle all considered hierarchies are compatible with the experimental bounds on the neutrino mixing angles in our model, the statistics summarized in the following support the previous statement: 
\begin{align}
\label{e:statnh} &&&&&&&&&&\text{NH:}\qquad &m_0 \in \left[0.00, 0.05\right]\,\text{eV},\qquad &n=46250,\qquad &n_L=19,&&&&&&&&&& \\
\label{e:statih} &&&&&&&&&&\text{IH:}\qquad &m_0 \in \left[0.00, 0.05\right]\,\text{eV},\qquad &n=3768,\qquad &n_L=229,&&&&&&&&&& \\
\label{e:statdeg} &&&&&&&&&&\text{deg:}\qquad &m_0 \in \left[0.05, 0.10\right]\,\text{eV},\qquad &n=189,\qquad &n_L=0,&&&&&&&&&&
\end{align}
where $n$ and $n_{\text{L}}$ count the number of events with and without leptogenesis, respectively, for each considered mass regime. The neutrino mixing is best accounted for by the NH scenario, whereas IH is favored if leptogenesis successfully explains the baryon asymmetry of the Universe. The RG analysis reveals that the model cannot accommodate the leptonic mixing angles for $m_0 > 0.056~\text{eV}$, practically ruling out degenerate neutrino masses as a suitable scenario.

Because of the number of combinations accommodating correct neutrino mixing as well as the baryon asymmetry of the early universe, predictions are the most reliable in the IH scenario. The predictions on the $CP$ phases listed in Table \ref{tab:lep} can be used for further studies; e.g., the Majorana phases $\alpha$ and $\beta$ affect the observables measured in $0\nu\beta\beta$ experiments. The amplitude of these processes is proportional to the effective Majorana mass
\begin{align}
m_{\beta\beta} = \sum_i U_{ei}^2\, m_i.
\end{align}
Cancellation can occur in NH depending on the size of the Majorana phases; however, in this particular case we predict small low energy $CP$ violation as can be seen from Table \ref{tab:lep}. According to our results above we give an estimate of the effective Majorana mass in the considered hierarchies, 
\begin{align}
 m_{\beta\beta}^{\text{NH}} \in \left[0.048, 0.063\right]\,\text{eV}, \qquad m_{\beta\beta}^{\text{IH}} \in \left[0.026, 0.050\right]\,\text{eV}.
\end{align}
Note again that due to Eqs. (\ref{e:statnh}) - (\ref{e:statdeg}) the predictions are most reliable in the IH scenario. The bounds on $m_{\beta\beta}$ are well below the current upper limit $\langle m_{\beta\beta}\rangle \lesssim 0.19-0.45~$eV given by the EXO-200 experiment at 90\% confidence level \cite{Albert:2014awa}, but partly accessible in next-generation $0\nu\beta\beta$ decay experiments.

\section{Conclusions}
\label{s:con}
$SU(5)$ inspired seesaw type I models with an almost diagonal Dirac sector and three heavy right-handed neutrinos successfully accommodate $\theta_{13}^{\text{Exp}}$. The large neutrino mixing angles are ascribed to the structure of the right-handed sector, while the small mixing angle is generated by a complex perturbation parameter $\epsilon$ and a real parameter $\gamma$ assigned to the off-diagonal elements of $m_D$. The structure of the perturbed Dirac mass matrix is obtained from an embedding in an $SU(5) \times U(1) \times \mathbb{Z}'_2 \times \mathbb{Z}''_2 \times \mathbb{Z}'''_2$ symmetry, where the hierarchy among the fermion families is generated by a Froggatt-Nielsen mechanism. The parameter $\epsilon$ also provides a possible source of $CP$ violation in the leptonic sector, enabling predictions on the Dirac $CP$ phase $\delta$ and the Majorana phases $\alpha$ and $\beta$. 

The numerical analysis shows that the neutrino mixing parameters can in principle be reproduced in all considered hierarchies for masses up to $m_0 \approx 0.056~\text{eV}$. However, the NH scenario is strongly favored over IH and the degenerate mass regime accommodating the correct neutrino mixing angles $\theta_{ij}$. Further constraints can arise from the requirement that leptogenesis generates the baryon asymmetry of the Universe, which eventually results in a preference of IH and rules out degenerate neutrino masses entirely. Regarding the phases $\delta$, $\alpha$, and $\beta$ the $CP$ violation in the IH case can be large, whereas for NH we predict small $CP$ violation in all phases. Since the degenerate mass regime is excluded, the resulting effective Majorana mass $m_{\beta\beta}$ is partly accessible by next-generation $0\nu\beta\beta$ experiments in IH scenarios. 

The leptogenesis scenario discussed in this paper is in good agreement with experimental and cosmological bounds, although a more realistic scenario with higher precision may improve the final predictions.  

\section{Acknowledgments}
H.P. was supported by DFG Grant No. PA 803/6-1. We thank Gautam Bhattacharyya for many helpful comments both on the original idea as 
well as on the actual work.

\nocite{*}
\bibliography{mybibpap2}
\bibliographystyle{h-physrev}

\end{document}